\documentclass[twocolumn]{aastex631}

\usepackage{graphicx}
\usepackage{amsmath}
\usepackage{float}
\usepackage{natbib}
\usepackage{tabularx}
\usepackage{bm}
\usepackage{appendix}
\usepackage{longtable}
\usepackage{url}

\newcommand\msun{M_{\odot}}

\shorttitle{Tidal Capture of WDs by IMBHs}
\shortauthors{Ye et al.}

\graphicspath{{./}{figures/}}

\begin{document}

\title{On the Tidal Capture of White Dwarfs by Intermediate-mass Black Holes in Dense Stellar Environments}

\author[0000-0001-9582-881X]{Claire S. Ye}
\affil{Canadian Institute for Theoretical Astrophysics, University of Toronto, 60 St. George Street, Toronto, Ontario M5S 3H8, Canada}
\correspondingauthor{Claire S.~Ye}
\email{claireshiye@cita.utoronto.ca}

\author[0000-0002-7330-027X]{Giacomo Fragione}
\affil{Department of Physics \& Astronomy, Northwestern University, Evanston, IL 60208, USA}
\affil{Center for Interdisciplinary Exploration \& Research in Astrophysics (CIERA), Northwestern University, Evanston, IL 60208, USA}

\author[0000-0002-3635-5677]{Rosalba Perna}
\affil{Department of Physics and Astronomy, Stony Brook University, Stony Brook, NY, 11794, USA}
\affil{Center for Computational Astrophysics, Flatiron Institute, 162 5th Avenue, New York, NY 10010, USA}

\begin{abstract}

Intermediate-mass black holes (IMBHs) are the missing link between stellar-mass and supermassive black holes, widely believed to reside in at least some dense star clusters, but not yet observed directly. Tidal disruptions of white dwarfs (WDs) are luminous only for black holes less massive than $\sim 10^5\,\msun$, therefore providing a unique smoking gun that could finally prove the existence of IMBHs beyond any reasonable doubt. Here, we investigate the tidal captures of WDs by IMBHs in dense star clusters, and estimate upper limits to the capture rates of $\sim 1\,{\rm Myr}^{-1}$ for galactic nuclei and $\sim 0.01\,{\rm Myr}^{-1}$ for globular clusters. Following the capture, the WD inspirals onto the IMBH producing gravitational waves detectable out to $\sim100$~Mpc by LISA for $\sim 10^4\msun$ IMBHs. The subsequent tidal stripping/disruption of the WD can also release bright X-ray and gamma-ray emission with luminosities of at least $\gtrsim10^{40}\,\rm{erg\,s^{-1}}$, detectable by \textit{Chandra}, \textit{Swift}, and upcoming telescopes, such as the \textit{Einstein Probe}.

\end{abstract}


\section{Introduction} \label{sec:intro}
Intermediate-mass black holes (IMBHs) are some of the most mysterious objects in the Universe. They are often introduced as the black holes (BHs) occupying the mass range between stellar-mass BHs ($\sim 10\,M_{\odot}$) and supermassive black holes (SMBHs; $10^6-10^{10}\,M_{\odot}$). IMBHs may play a crucial role in cosmology and galaxy formation, as they could act as building blocks of SMBHs that are observed at the centers of most galaxies \citep[see e.g.,][and references therein]{Greene+2020imbh}. Understanding how IMBH masses are related to their environments (e.g., velocity dispersion) may also provide unique insights into the dynamical evolution of dense star clusters \citep[e.g.,][]{Miller_Hamilton_2002,PZ_McMillan_2002,Gurkan+2004,SubrFragione2019}. Furthermore, the inspirals of stellar compact remnants into an IMBH could provide an extraordinarily powerful tool for testing general relativity in the strong field regime \citep[e.g.,][]{Miller_2002}. Despite these rich physical applications, the existence of IMBHs has long been debated and very little is known about their origin and evolution. Only recently was their existence confirmed, when a $150\,M_{\odot}$ binary BH merger was detected by LIGO/Virgo \citep{gw190521}.

IMBHs are most likely produced in dense stellar environments, such as globular clusters (GCs) and galactic nuclei. They could form through collisional runaways of massive stars \citep[e.g.,][]{PZ_McMillan_2002,Giersz+2015,Kremer+2020imbh,Gonzalez+2021} or through repeated mergers of stellar-mass BHs \citep[e.g.,][]{Miller_Hamilton_2002,AntoniniGieles2019,FragioneKocsis2022,MapelliBouffanais2022}. Here, IMBHs can also have frequent dynamical encounters with stars and compact objects. For example, during these interactions, it is very likely that the IMBHs will quickly form binaries with other compact objects through exchange encounters. The subsequent intermediate-mass inspirals of the stellar remnants into the IMBHs can produce gravitational wave (GW) sources detectable by LISA and ground-based telescopes such as the Einstein Telescope \citep{JaniShoemaker2020,FragioneLoeb2023}. 

IMBHs are also uniquely suited at tidally disrupting white dwarfs (WDs). The tidal disruption radius of a WD is outside the event horizon of an IMBH; on the other hand, the tidal disruption radius of a WD is inside the event horizon of an SMBH, so no electromagnetic waves can be observed, and a stellar-mass BH will enter a WD instead of disrupting it during close encounters \citep{Luminet_Pichon_1989,Maguire+2020}. Since WD tidal disruptions are a smoking gun for revealing IMBHs, they have been under intense scrutiny \citep[e.g.,][]{Baumgardt+2004,Rosswog+2008a,Rosswog+2008b,Sesana+2008,Rosswog+2009,Zalamea+2010,Clausen_Eracleous_2011,Haas+2012,MacLeod+2014,MacLeod+2016_optical,MacLeod+2016_close,Tanikawa+2017,Anninos+2018,Fragione+2018tde,Fragione_Leigh_2018imri,Kawana+2018,Toscani+2020,Chen+2022,Tanikawa+2022}.

Despite the wealth of studies on WD tidal disruptions, little attention has been devoted to computing the tidal capture rates of WDs, with the possible exception of \citet{Ivanov_Papaloizou_2007}. In this study, we explore in detail the characteristics of IMBH-WD tidal interactions in both galactic nuclei and GCs. During close encounters, the tidal force exerted by an IMBH can induce oscillations within the WD, which can then lead to the formation of an IMBH-WD binary. This mechanism for capturing WDs is different from binary splitting according to the Hills mechanism which requires a binary or the scattering of WDs into the loss cone, and thus provides an additional channel for IMBH-WD binary formation \citep{Hills_1988}. These captured binaries are prime sources of low-frequency gravitational radiation during their inspiral and will provide key GW sources for future space detectors such as LISA \citep[e.g.,][]{Amaro-Seoane+2007,Amaro-Seoane+2017,Fragione+2018imbh}. Subsequent tidal stripping/disruption of the WD can be observed by ongoing and future time-domain surveys \citep[e.g.,][for a review]{Maguire+2020}. These multi-wavelength detections will provide definitive evidence for the existence of IMBHs. 

Our paper is organized as follows. In Section~\ref{sec:methods}, we describe the analytical methods used to calculate the WD oscillation energy during close passages to the IMBHs. In Section~\ref{sec:results}, we discuss the WD tidal capture rates in different dense star clusters, while we show the potential multi-wavelength signals from gravitational radiation and WD stripping/disruption in Section~\ref{sec:observ}. Lastly, we discuss the uncertainties and conclude in Section~\ref{sec:conclu}.

\section{Tidal Captures of White Dwarfs} \label{sec:methods}
In this Section, we describe the methods used for estimating the tidal capture rates of WDs by an IMBH in galactic nuclei and GCs.

Throughout this study, we fix the mass of the WDs to be $0.6~\msun$ and adopt Eq.~91 in \citet{SSE} for the radius of a WD. These low-mass WDs are well represented by an $n=1.5$ polytrope \citep{Shapiro_Teukolsky_1983}. Following \citet{Fabian+1975}, \citet{Press_Teukolsky_1977}, and \citet{Lee_Ostriker_1986}, we estimate the amount of oscillation energy deposited into a WD that is passing at a distance $R_{p}$ from an IMBH as

\begin{equation}\label{eq:eosc}
    \Delta E_{osc} = \frac{GM_{\rm{WD}}^2}{R_{\rm{WD}}} \left(\frac{M_{\rm{IMBH}}}{M_{\rm{WD}} }\right)^2 \sum^\infty_{l=2}\left(\frac{R_{\rm{WD}}}{R_{p}}\right)^{2l+2} T_l(\eta)\,,
\end{equation}
where $M_{\rm{WD}}$ and $R_{\rm{WD}}$ are the mass and radius of the WD, respectively, $T_l$ is a dimensionless function measuring the contributions from different harmonic modes $l$ (we include the quadruple and octupole terms), and $\eta$ is the duration of the periastron passage

\begin{equation}\label{eq:eta}
    \eta = \left(\frac{M_{\rm{WD}}}{M_{\rm{WD}}+M_{\rm{IMBH}}}\right)^{1/2} \left(\frac{R_p}{R_{\rm{WD}}}\right)^{3/2}\,.
\end{equation}
This linear energy approximation agrees well with more detailed energy calculations except at small $R_p$ \citep[e.g.,][and in particular their Figure~9]{Cheng_Evans_2013}.

The maximum tidal capture radius, $R_{cap}$, is obtained when the total initial kinetic energy of the two interacting objects, $E_{orb}=\frac{1}{2}\mu \sigma_v^2$ ($\sigma_v$ is the relative velocity of the IMBH and WD), equals $\Delta E_{osc}$, assuming fast dissipation of the oscillation energy. 

\begin{figure}
\begin{center}
\includegraphics[width=\columnwidth]{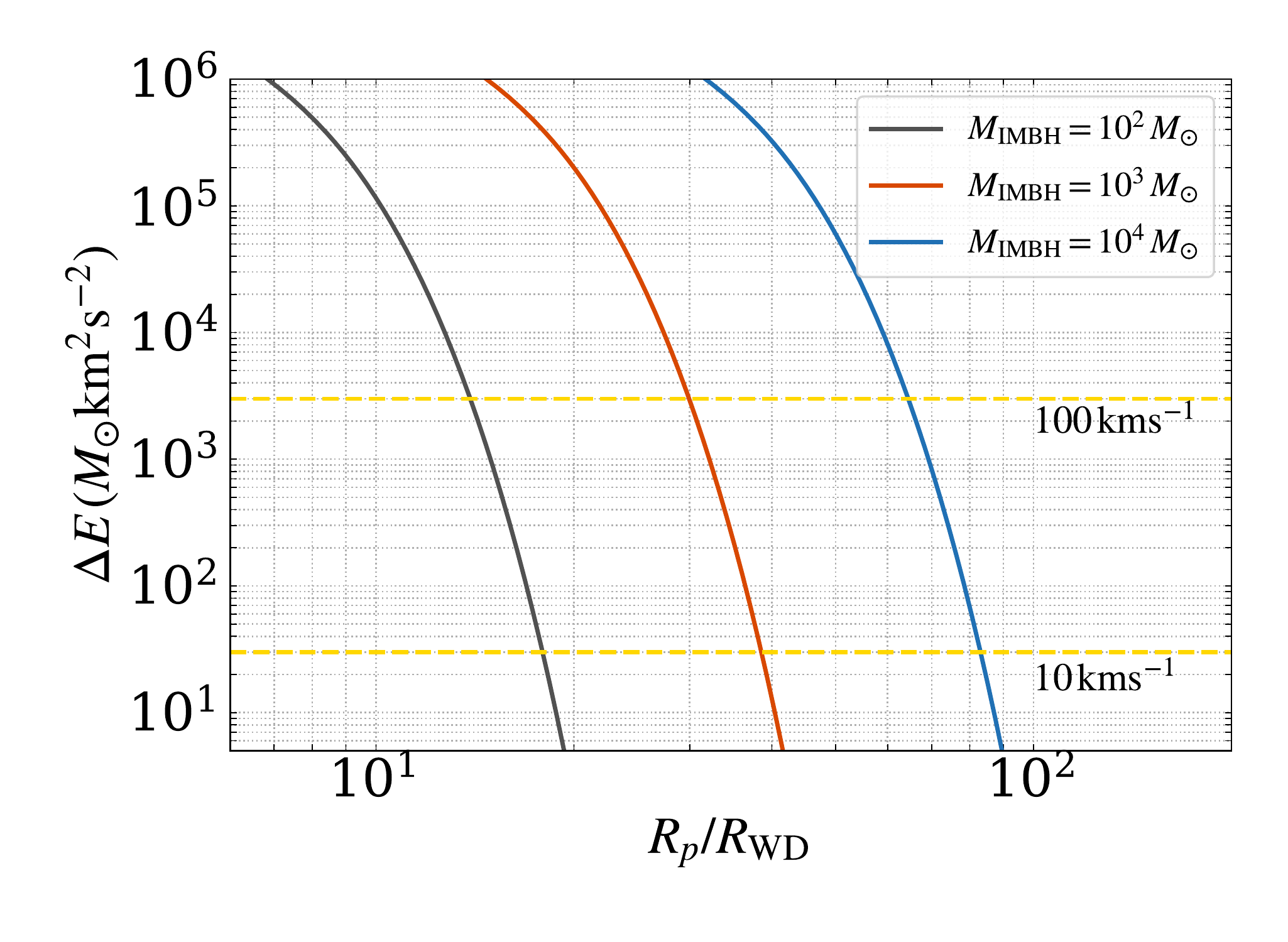}
\caption{The amount of energy deposited into a WD after one periastron passage as a function of the pericenter distance in the unit of the WD radius. The two yellow horizontal lines show the total initial kinetic energy of a $10^3\,\msun$ IMBH and a $0.6\,\msun$ WD with relative velocities at infinity of $10\,\rm{km\,s^{-1}}$ and $100\,\rm{km\,s^{-1}}$, respectively. For small pericenter distances, the oscillation energy is larger than the total kinetic energy.}\label{fig:Eosc}
\end{center}
\end{figure}

The effect of an IMBH's tidal force on a WD is stronger the closer the WD is to the IMBH as can be seen from Eq.~\ref{eq:eosc}. Figure~\ref{fig:Eosc} shows the amount of orbital energy injected into a WD during the first passage of tidal interactions with IMBHs of different masses. For example, for a $10^3\,\msun$ IMBH and WD encounter where the pericenter distance is smaller than $\sim 30 R_{\rm{WD}}$, the energy deposited into the WD and subsequently dissipated through its oscillation exceeds the total initial kinetic energy assuming a velocity dispersion of $100\,\rm{km\,s^{-1}}$. For lower velocity dispersions, e.g., in GCs where $\sigma_v$ is $\sim 10\,\rm{km\,s^{-1}}$, the minimum pericenter distance is larger, at $\sim40R_{\rm{WD}}$. For all IMBH masses, the maximum capture radii in GCs are about $1.3$ times larger than in galactic nuclei for the velocity dispersions considered in Figure~\ref{fig:Eosc}.

The capture cross section can then be written as \citep{Quinlan_Shapiro_1987}
\begin{equation}
    \sigma_{cap} = \pi R_{cap}^2\left(1+\frac{2GM_{\rm{IMBH}}(1+q)}{R_{cap} v_{rel}^2}\right),
\end{equation}
where $q = M_{\rm{WD}}/M_{\rm{IMBH}}$ and $v_{rel}$ is the relative velocity between the IMBH and the WD. We adopt in the calculations $v_{rel}\approx \sigma_v$ as typical, where $\sigma_v$ is the local velocity dispersion, but note that $v_{rel}$ can be up to $2\sigma_v$. Thus, the rates of IMBH-WD tidal captures (per IMBH) can be estimated by the two-body collision rates
\begin{equation}\label{eq:coll}
   \Gamma = n \sigma_{cap} \sigma_v\,,
\end{equation}
where $n$ is the local WD number density.\footnote{We assume that the loss cone of the IMBH is full and continuously replenished by WDs since the IMBH is moving through the surrounding cusp and the WDs would have enough time to refill the loss cone. If the loss cone becomes empty, fewer capture/disruption events would occur.} The timescale for tidal capture is
\begin{equation}\label{eq:t_coll}
    T_{\rm{IMBH-WD}} = \Gamma^{-1}\,.
\end{equation}

At the moment of tidal capture, the semi-major axis of the newly-formed IMBH-WD binary, $a$, can be calculated from its binding energy $E_b$
\begin{equation}\label{eq:sma} 
    a = \frac{G M_{\rm{IMBH}} M_{\rm{WD}}}{2 |E_b|}\,,
\end{equation} and the eccentricity, $e$, can then be derived from $R_{cap} = a(1-e)$. For $E_b$ close to zero, the orbit will tend to be parabolic, and $a$ and $e$ will be close to infinity and unity, respectively. To estimate the minimum $a$ and $e$, we assume that the total initial energy, $E_{orb}$, is zero, and that $E_b = \Delta E_{osc}$ at the WD's tidal disruption radius. We define the tidal disruption radius of a WD to be \citep[e.g.,][]{Rees_1988}
\begin{equation}
    R_{t} = \left(\frac{M_{\rm{IMBH}}}{M_{\rm{WD}}}\right)^{1/3} R_{\rm{WD}}\,.
    \label{eq:Rt}
\end{equation}
The radius of a WD with $M_{\rm{WD}}=0.6\,\msun$ is about $0.013\,R_{\odot}$. For IMBHs with masses of $10^2$, $10^3$, and $10^4\,\msun$, the tidal disruption radius of a WD is $5.5\,R_{\rm{WD}}$ ($3\times10^{-4}$~AU), $11.9\,R_{\rm{WD}}$ ($7\times10^{-4}$~AU), and $25.5\,R_{\rm{WD}}$ ($1.5\times10^{-3}$~AU), respectively. The minimum capture semi-major axis and eccentricity, $[a, e]$, are then estimated to be $[0.012\,\rm{AU}, 0.973]$, $[0.12\,\rm{AU}, 0.994]$, and $[1.24\,\rm{AU}, 0.999]$, respectively. Note that these are approximate estimates since the oscillation energy in Eq.~\ref{eq:eosc} may have non-negligible contributions from higher-order terms when $R_p\sim R_t$. Both the semi-major axis and eccentricity are larger for binaries captured at larger pericenter distances than the tidal disruption radii (Eq.~\ref{eq:sma}). The maximum capture semi-major axes are about $160$, $1641$, and $16456$~AU for IMBHs with masses of $10^2$, $10^3$, and $10^4\,\msun$, respectively, for captures at about $3R_t$, which is about the maximum tidal capture radius of the WD (Figure~\ref{fig:rcap_nsc} and \ref{fig:rcap_gc}). At the same time, the eccentricities tend to be unity (see also Figure~\ref{fig:ecc} below).

\section{Rates} \label{sec:results}
In this Section, we analyze the capture rates for IMBHs of masses $10^2$, $10^3$, and $10^4\,\msun$ in different dense stellar environments.

\subsection{Galactic Nucleus Environments}\label{subsec:nsc}
Here, we consider various galactic nucleus environments hosting SMBHs with masses $M_{\rm{SMBH}}=10^5, 10^6$, and $10^7~\msun$. The radius of influence of an SMBH is defined as \citep{Merrit_2013}
\begin{equation}\label{eq:rh}
    R_h = \frac{GM_{\rm{SMBH}}}{v_h^2}\,,
\end{equation}
where $v_h$ is the velocity dispersion at $R_h$. The velocity dispersion is a function of the galactocentric distance $r$, $\sigma_v \sim (G M_{\rm{SMBH}}/r)^{1/2}$. Following \citet{Tremaine+2002} for the `$\rm{M-\sigma}$' relation, $v_h$ can be expressed as a function of the mass of the SMBH
\begin{equation}\label{eq:msig}
    M_{\rm{SMBH}} = \gamma v_h^4\,,
\end{equation}
and $\gamma \approx 0.0758\,\msun\,\rm{(km\,s^{-1})^{-4}}$ \citep[][Eq. 19]{Tremaine+2002}. Combining Eq.~\ref{eq:rh} and Eq.~\ref{eq:msig}, we can write the radius of influence as 
\begin{equation}
    R_h = G\gamma^{1/2}M_{\rm{SMBH}}^{1/2}\,.
\end{equation}

We assume power-law number density distributions for the main-sequence stars and WDs surrounding the SMBHs \citep[][and references therein]{Gondan+2018}. The distribution function can be written as
\begin{equation}
    n_{\rm{WD}} = n_{inf}\left(\frac{r}{R_h}\right)^{-\alpha}\,,
\end{equation}
where
\begin{equation}
    n_{inf} = 1.38\times10^4 \sqrt{\frac{10^6\,\msun}{M_{\rm{SMBH}}}}\,{\rm pc}^{-3}\,.
\end{equation}
The number density of main-sequence stars is $10$~times larger following the same distribution.

In galactic nuclei, a massive binary will gradually inspiral through dynamical friction. Following \citet[][Eq. A8]{Gurkan_Rasio_2005}, and using the definition that the enclosed mass within $R_h$ equals $2M_{\rm{SMBH}}$, we can write the dynamical friction time at a distance $r$ as
\begin{equation}
    \begin{aligned}\label{eq:tdf}
     T_{df} &= \frac{[1+(1+\beta)(r/R_h)^\beta/2][1+(r/R_h)^\beta]^{0.5}(r/R_h)^{1.5}}{1.5\chi \rm{log\Lambda} \beta (r/R_h)^\beta}\\
    &\times Q T_h\approx 2.67\,\rm{Myr}\,\left(\frac{Q}{1000}\right)\\
    &\times \frac{[1+(1+\beta)(r/R_h)^\beta/2][1+(r/R_h)^\beta]^{0.5}(r/R_h)^{1.5}}{\chi \beta (r/R_h)^\beta}\,,
    \end{aligned}
\end{equation}
where $\beta = 3-\alpha$, $\rm{log \Lambda} \sim 5$ is the Coulomb logarithm \citep{McMillan_PZ_2003}, $Q = M_{\rm{SMBH}}/M_{\rm{IMBH}}$, the factor $\chi$ can be written as \citep{BT_2008}
\begin{equation}
    \chi = erf(X)-\frac{2X}{\sqrt{\pi}}e^{-X^2}
\end{equation}
with $X = \sqrt{2-\beta}$ \citep{McMillan_PZ_2003}, and $T_h=R_h/v_h$. The dynamical friction timescale at the radius of influence of a $4\times10^6\msun$ SMBH is about $2000$, $200$, and $20$~Myr for $M_{\rm{IMBH}}=10^2$, $10^3$ and $10^4~\msun$, respectively. For these values, the power-law slope of the stellar distribution is assumed to be $\alpha = 1.4$, as derived in numerical calculations of two-body relaxation \citep{Hopman_Alexander_2006}.

For shorter distances to the SMBHs, the inspiral of the massive binary is dominated by GW radiation instead, and the GW inspiral timescale is \citep{Peters_1964}
\begin{equation}\label{eq:tgw}
    \begin{aligned}
        T_{\rm{GW}} &= \frac{5}{256} \frac{a_{\rm{IMBH}}^4 c^5}{G^3 M_{\rm{IMBH}}^3 Q(1+Q)} (1-e_{\rm{IMBH}}^2)^{7/2}\\
        &\approx 5.84\times10^4\,\rm{Gyr}\frac{1}{Q(1+Q)}\left(\frac{a_{\rm{IMBH}}}{10^{-4}\rm{pc}}\right)^4\left(\frac{M_{\rm{IMBH}}}{10^3\msun}\right)^{-3},
    \end{aligned}
\end{equation}
where $c$ is the speed of light, and $a_{\rm{IMBH}}$ and $e_{\rm{IMBH}}$ are the semi-major axis and eccentricity of the IMBH orbiting the SMBH, respectively. The inspiral timescale due to GW emission is about $10^7$, $10^6$, and $10^5$~Myr for $M_{\rm{IMBH}}=10^2$, $10^3$, and $10^4~\msun$, respectively, assuming the IMBHs are orbiting a $4\times10^6\msun$ SMBH at $10^{-3}$ times its radius of influence on a circular orbit.

\begin{figure}
\begin{center}
\includegraphics[width=\columnwidth]{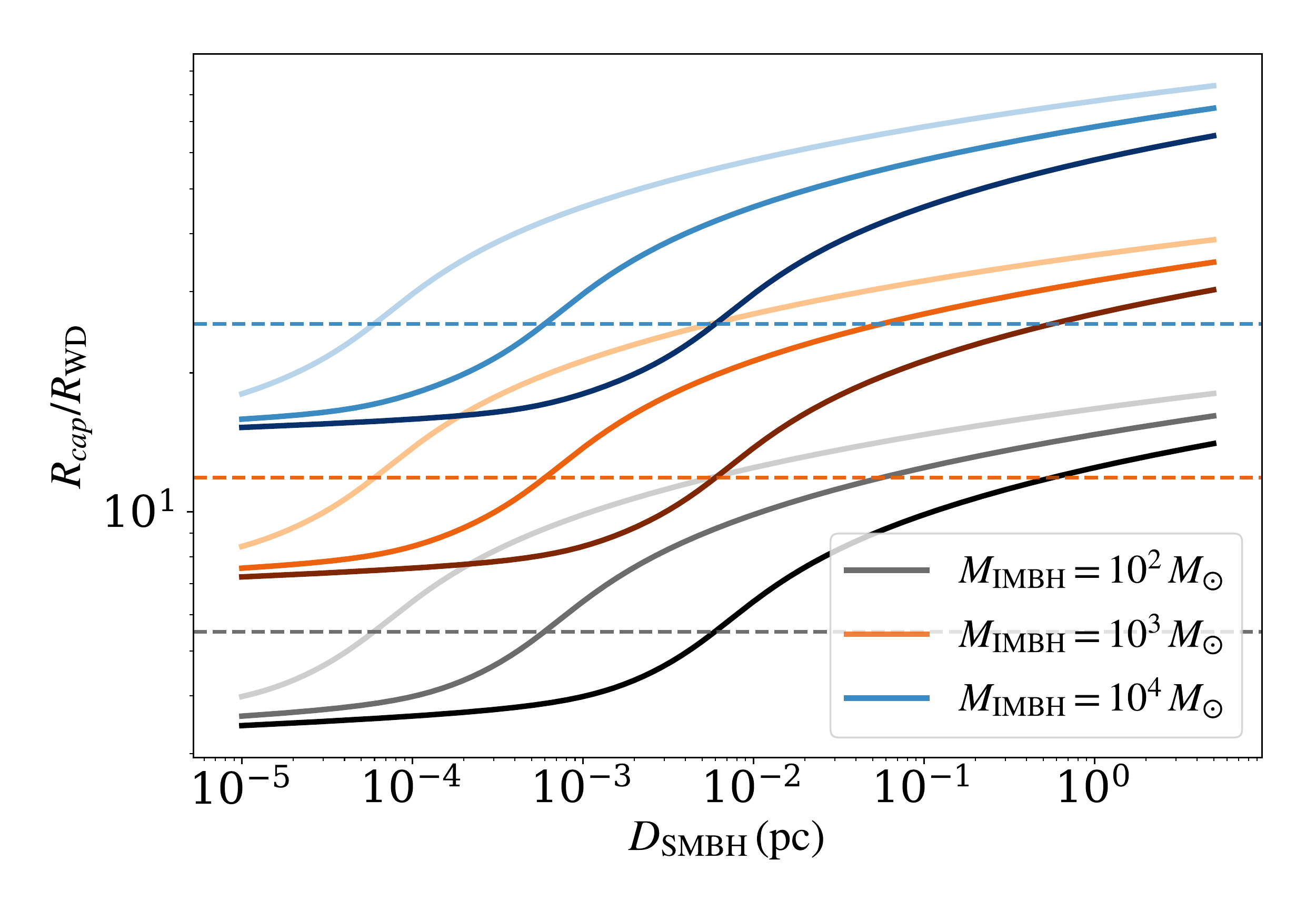}
\caption{Maximum tidal capture radii for various IMBH and SMBH masses as a function of the distances to the SMBHs in galactic nuclei. The three colors/groups of curves correspond to three IMBH masses of $10^4$ (blue), $10^3$ (orange), and $10^2~\msun$ (black) from top to bottom. The curves in each group from top to bottom correspond to increasing SMBH masses of $10^5$, $10^6$, and $10^7\,\msun$, respectively. The horizontal lines show the tidal disruption radii for different IMBH masses.}\label{fig:rcap_nsc}
\end{center}
\end{figure}

We show in Figure~\ref{fig:rcap_nsc} the maximum tidal capture radius in galactic nuclei. The maximum capture radius increases as the distance to the center of galaxies increases or the mass of the central SMBH decreases, which causes the velocity dispersion to decrease. For very small galactocentric distances, the maximum tidal capture radii could be smaller than the tidal disruption radii. At these distances, no tidal capture binaries can form during close encounters between IMBHs and WDs, and the IMBHs will directly disrupt the WDs.

\begin{figure*}
\begin{center}
\includegraphics[width=\textwidth]{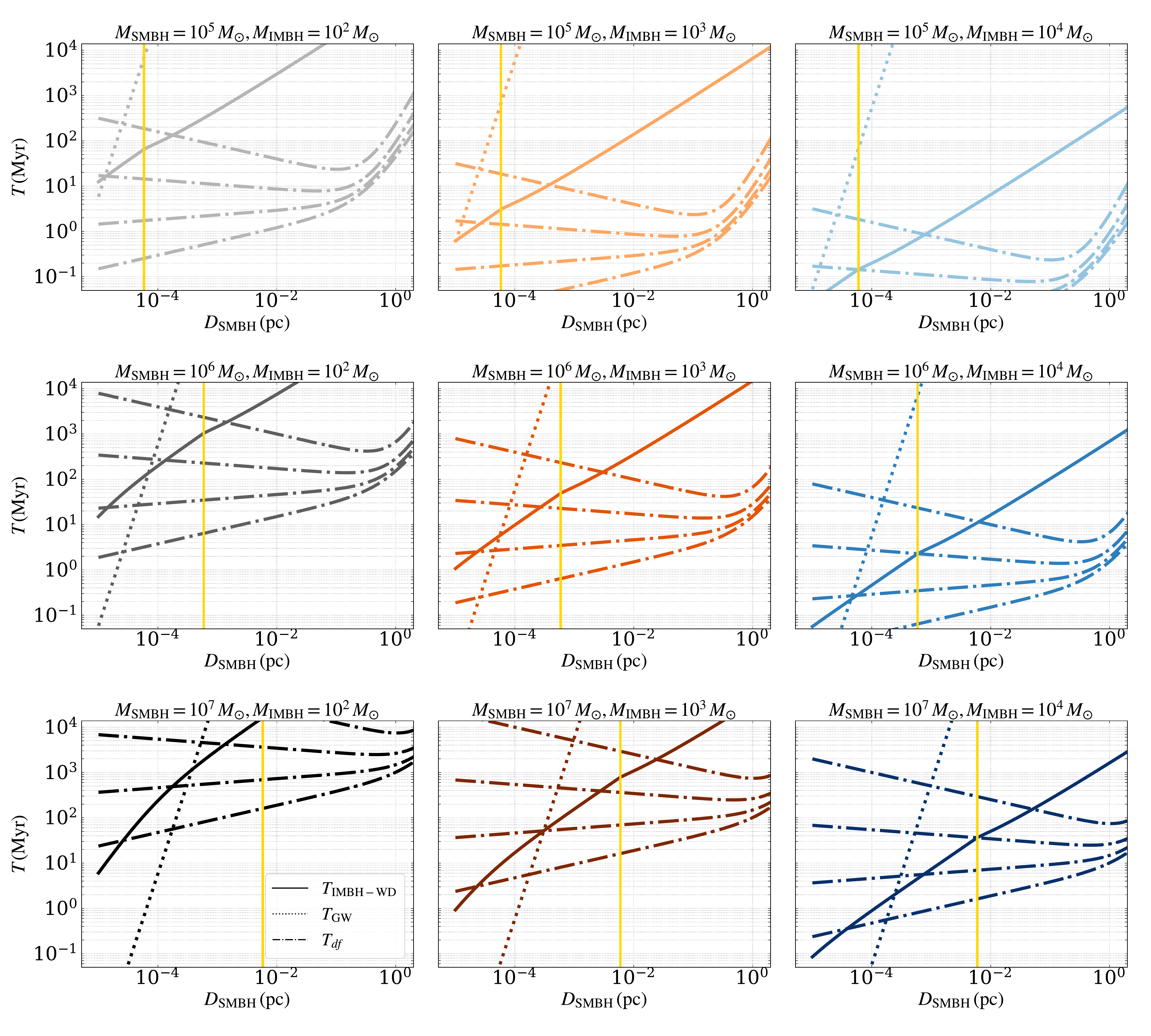}
\caption{Comparison between timescales. In each panel, the solid curve shows the tidal capture/disruption timescale ($T_{\rm{IMBH-WD}}$; Eq.~\ref{eq:t_coll}) at a given galactocentric distance. The vertical yellow line is the distance at which the maximum tidal capture radius is smaller than the tidal disruption radius. On the right side of the yellow line, the solid curve shows the tidal capture timescale, and on the left shows the timescale of the tidal disruption events (TDEs). The dotted line shows the GW inspiral timescale of the IMBH towards the SMBH (Eq.~\ref{eq:tgw}; assuming $e_{\rm{IMBH}}=0$). The dotted-dashed curves indicate the dynamical friction timescales for stellar number density distributions with different slopes $\alpha$. From top to bottom, $\alpha=1.2$, 1.4, 1.6, and 1.8, respectively. The slope of the WD number density distribution is fixed to be $1.4$ \citep{Hopman_Alexander_2006} for the calculation of the tidal capture/disruption timescales. Tidal capture interactions or disruptions are possible when $T_{\rm{IMBH-WD}}<\rm{min}(\mathit{T_{df}, T_{\rm{GW}}})$.} \label{fig:times}
\end{center}
\end{figure*}

In addition, a WD can only be captured in the galactic nuclei before the IMBHs spiral into the central SMBHs. A successful capture depends on the dynamical friction (Eq.~\ref{eq:tdf}) and GW inspiral (Eq.~\ref{eq:tgw}) timescales of the IMBH, and needs to satisfy the condition $T_{\rm{IMBH-WD}}<\rm{min}(T_{\mathit{df}}, T_{\rm{GW}})$. We compare in Figure~\ref{fig:times} the typical timescale for an IMBH to capture or tidal disrupt a WD in galactic nuclei with the IMBH's dynamical friction and GW inspiral timescales. The capture/disruption timescale is the shortest for the most massive IMBH orbiting the least massive SMBH. For IMBHs with $M_{\rm{IMBH}}\gtrsim 10^3\,\msun$, tidal capture of a WD is possible for relatively flat stellar number density slopes $\alpha \lesssim 1.4$. Observations and theoretical studies have shown that the power-law slope of the stellar density distribution in the Milky Way nuclear star cluster is $\sim 1.4$ \citep{Gallego-Cano+2018I,Gallego-Cano+2020}, and may be as low as $\sim 1$ for main-sequence stars \citep{Schodel+2018II,Baumgardt+2018III}. Thus our Milky Way center can potentially produce tidally captured IMBH-WD binaries. For distances closer to the central SMBH, the WD capture radius is smaller than the tidal disruption radius so a WD cannot be captured (left of the yellow line). In this region when $T_{\rm{IMBH-WD}}<\rm{min}(\mathit{T_{\mathit{df}}, T_{\rm{GW}}})$, an IMBH will directly disrupt a WD during close encounters.

\begin{figure}
\begin{center}
\includegraphics[width=\columnwidth]{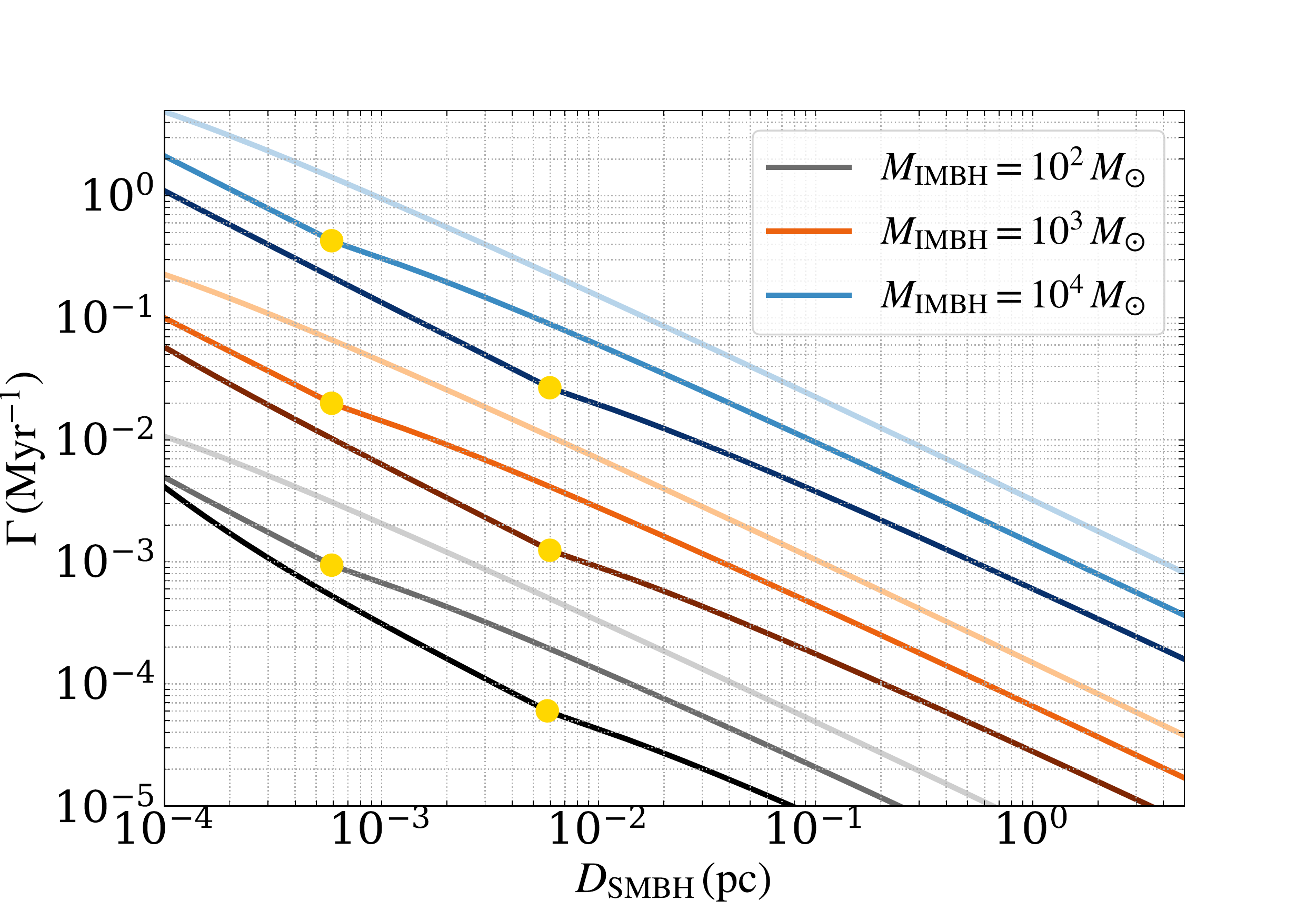}
\caption{The rates of WDs captured or disrupted tidally by IMBHs in the galactic nuclei as a function of the galactocentric distances. Each curve shows the rate of one IMBH and the colors have the same meanings as in Figure~\ref{fig:rcap_nsc}. The yellow dot on each curve marks the turning point where the maximum tidal capture radius becomes smaller than the tidal disruption radius. If a yellow dot is missing then for the range of distances shown in the figure the tidal capture radius is always larger than the tidal disruption radius.}\label{fig:rate_nsc}
\end{center}
\end{figure}

The capture and disruption rates as a function of the distance to the galactic center are shown in Figure~\ref{fig:rate_nsc}. The maximum tidal capture rate in a galactic nucleus is $\sim 5$ per Myr for an IMBH with $M_{\rm{IMBH}}=10^4\,\msun$ orbiting a $10^5\,\msun$ SMBH. For a more typical IMBH mass of $10^3\,\msun$ and an SMBH mass of $10^6\,\msun$, the tidal capture rate is $\sim 0.02$ per Myr. The tidal disruption rate in this case is $\sim 0.1$ per Myr.

\subsection{Globular Cluster Environments}\label{subsec:gc}
We assume that GCs can be modeled
with a Plummer density profile \citep{Plummer_1911, BT_2008}
\begin{equation}\label{eq:plummer}
    \rho(r) = \frac{3M_{cl}}{4\pi b^3}\left(1+\frac{r^2}{b^2}\right)^{-\frac{5}{2}},
\end{equation}
where $M_{cl}$ is the total mass of the cluster and $b$ is the Plummer scale length. The half-mass radius of a Plummer sphere is $r_{hm} \approx 1.3b$. We adopt a Kroupa initial mass function \citep{Kroupa_2001} between masses $0.08\,\msun$ and $180\,\msun$ and follow an initial-to-final mass relation from \citet[][Eq.~7.22]{Merrit_2013}. At the present day, about $17\%$ of the objects are WDs and we assume that the spatial distribution of the WDs follows the Plummer profile.

The velocity dispersion of a Plummer sphere is 
\begin{equation}
    \sigma_p^2(r) = \frac{G M_{cl}}{6\sqrt{r^2+b^2}}.
\end{equation}

Here, we assume that the IMBHs reside at the cluster centers. This is a reasonable assumption given that IMBHs are most likely formed from collisional runaways of stars \citep[e.g.,][]{PZ_McMillan_2002,Giersz+2015,Kremer+2020imbh,Gonzalez+2021} or repeated mergers of stellar-mass BHs \citep[e.g.,][]{Miller_Hamilton_2002,AntoniniGieles2019,FragioneKocsis2022,MapelliBouffanais2022}, which occur at around the centers of GCs. Besides, a massive IMBH would quickly mass segregate back to the cluster center in $\sim$Myr timescale if it is displaced by dynamical encounters. The least massive IMBHs may be more easily displaced or even ejected from their host clusters \citep[e.g.,][]{Gonzalez+2022}, but for this first rate estimate we ignore these effects.

\begin{figure}
\begin{center}
\includegraphics[width=\columnwidth]{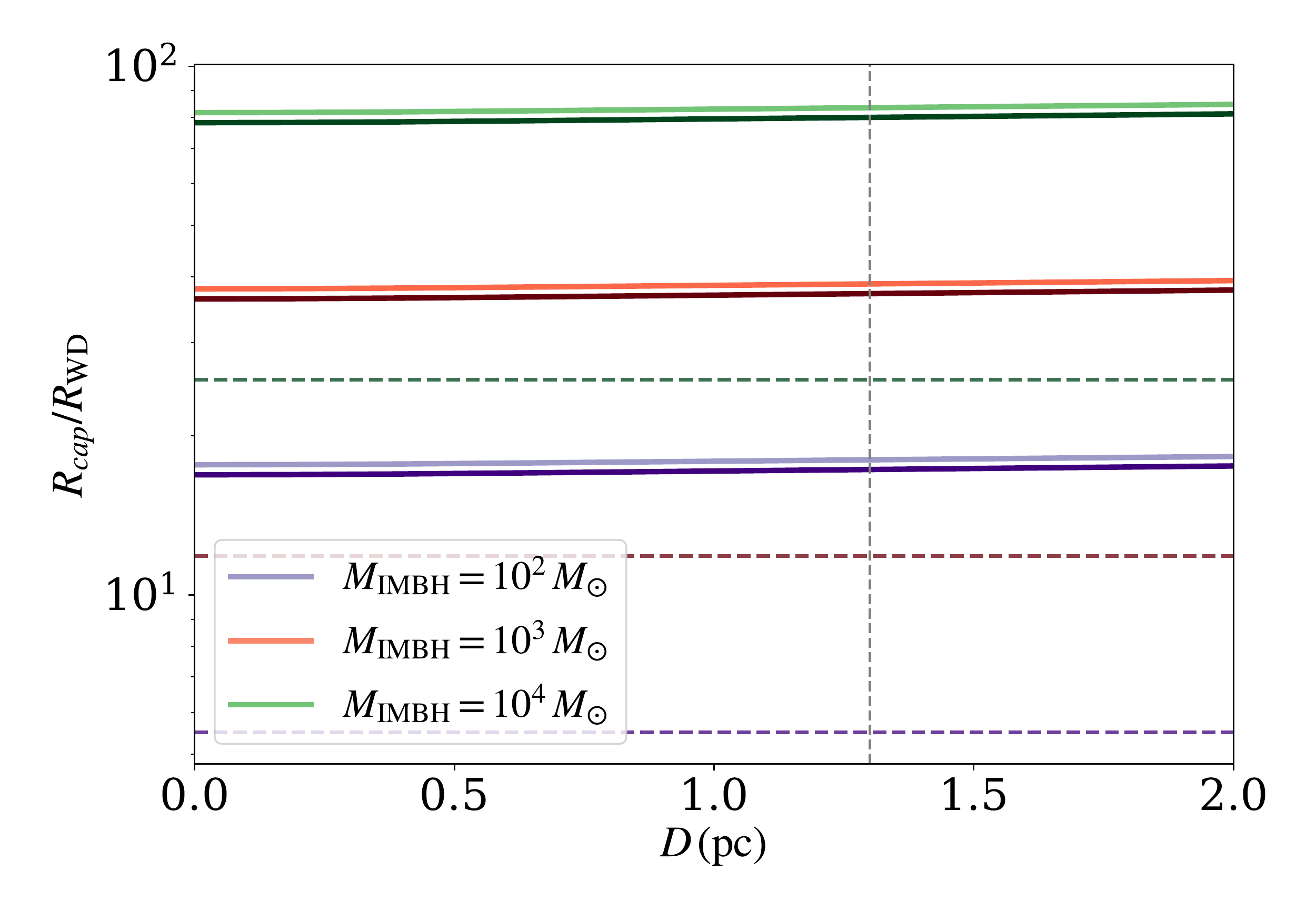}
\caption{Similar to Figure~\ref{fig:rcap_nsc}, but for GCs. The three colors/groups of curves correspond to three IMBH masses of $10^4$ (green), $10^3$ (red), and $10^2~\msun$ (purple) from top to bottom. For each group of curves, the top and bottom curves are for cluster masses of $2 \times 10^5$ and $5 \times 10^5\,\msun$, respectively. The vertical line shows the half-mass radius of the clusters where $b=1~$pc (Eq.~\ref{eq:plummer}).}\label{fig:rcap_gc}
\end{center}
\end{figure}

Similar to the galactic nuclei case in the above Section, the maximum tidal capture radius increases as the distance to the center of GCs increases or the total mass of GCs decreases, as is shown in Figure~\ref{fig:rcap_gc} (the increase is slow here). However, the maximum tidal capture radii in this case are always larger than the tidal disruption radii because the velocity dispersion in GCs is much smaller than in galactic nuclei. The tidal capture radius is about $3R_t$ for all IMBH masses.

\begin{figure}
\begin{center}
\includegraphics[width=\columnwidth]{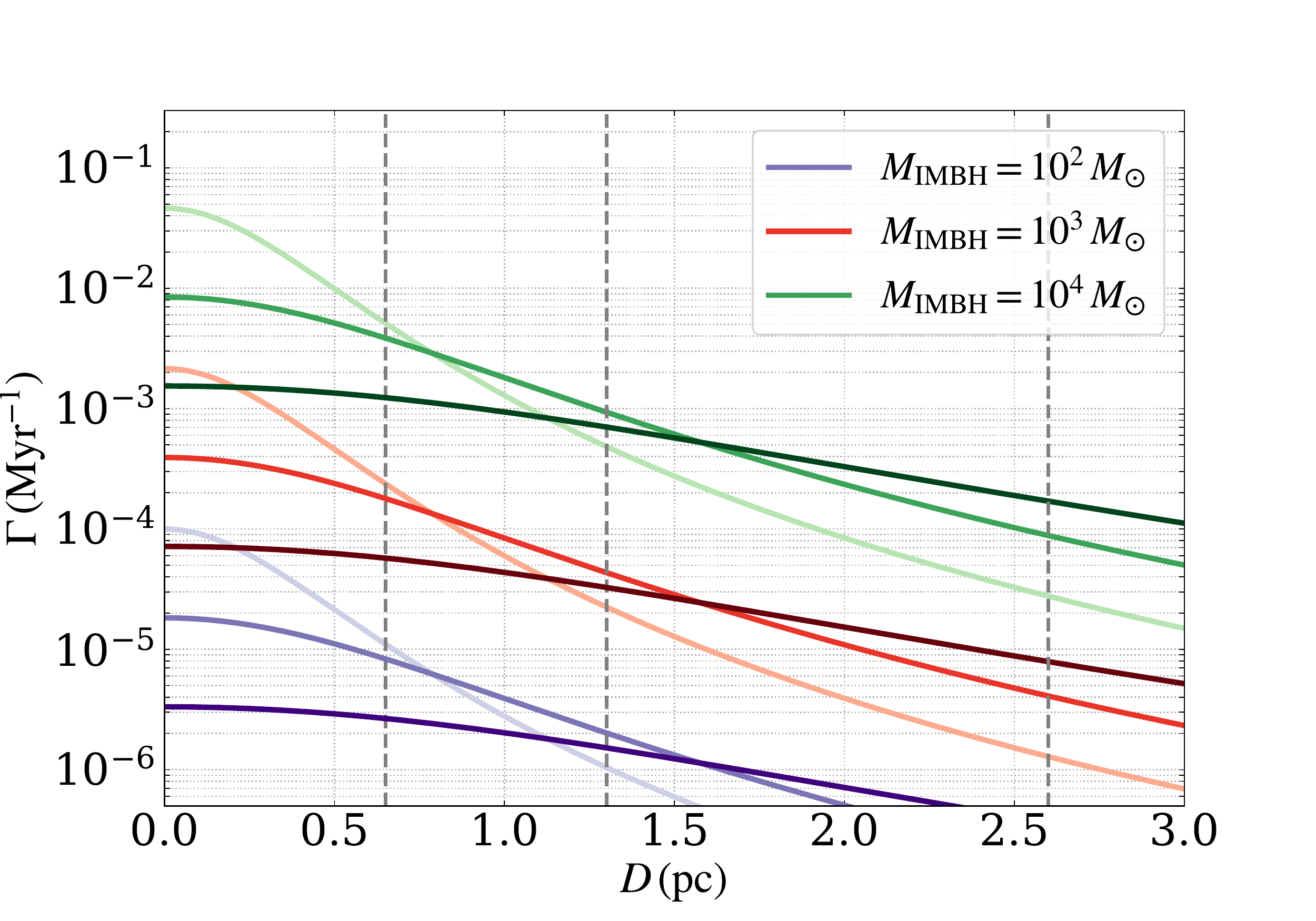}
\caption{Similar to Figure~\ref{fig:rate_nsc}, the rates of IMBH-WD tidal captures in GCs as a function of the distances to the cluster centers. Here we only show clusters with $M_{cl} = 2 \times 10^5\,\msun$, and each cluster has one IMBH. Different colors are for different IMBH masses, and each group of curves with the same color but different shades shows the Plummer scale length $b=0.5$, $1$, and $2~$pc from top to bottom. Smaller values of $b$ indicate denser clusters. The three vertical lines are the half-mass radii for these three scale lengths.}\label{fig:rates_gc}
\end{center}
\end{figure}

The capture rates as a function of the distance to the GC centers are shown in Figure~\ref{fig:rates_gc}. For a typical GC with $M_{cl} = 2\times10^5\,\msun$ and hosting a $10^3\,\msun$ IMBH at the center, the maximum tidal capture rate is $\sim 0.002$ per Myr. The tidal capture rates shown in Figure~\ref{fig:rates_gc} are also consistent with the rates calculated by \citet{Ivanov_Papaloizou_2007}. For a Milky Way-like galaxy hosting $\sim 200$ GCs, if we assume that all clusters have an IMBH, the optimistic tidal capture rate is a few times $0.1$ per Myr per galaxy, roughly comparable to the tidal capture rate from one IMBH in a galactic nucleus.

\section{Observational Signatures}\label{sec:observ}
We discuss here the multi-messenger observational signatures of these IMBH-WD interactions and the detectors that are sensitive to these signals.

\subsection{Gravitational Wave Emission}\label{subsec:gw}
After an IMBH-WD binary is formed, the two objects inspiral via the emission of GW radiation, and the emission will be observable by LISA. To estimate the amount of energy radiated by the binary, we compute the evolution of its semi-major axis and eccentricity by integrating Eq.~5.8 from \citet{Peters_1964} and stop the evolution when the pericenter distance $R_p = R_t$.

For eccentric binaries, the frequency at peak GW emission is given by \citep{Wen_2003}
\begin{equation}
    f_{peak} = \frac{\sqrt{G(M_{\rm{IMBH}}+M_{\rm{WD}})}}{\pi}\frac{(1+e)^{1.1954}}{[a(1-e^2)]^{1.5}}.
\end{equation}
The characteristic strain at the nth harmonic can be calculated as \citep{Barack_Cutler_2004}
\begin{equation}
    h_{c,n}^2=\frac{1}{(\pi D)^2}\left(\frac{2G}{c^3}\frac{\dot E_n}{\dot f_n}\right),
\end{equation}
where $D$ is the luminosity distance to the source, $\dot f_n$ is the time derivative of the GW frequency of the nth harmonic at the rest frame, and $\dot E_n$ is the GW power radiated at $f_n$. The frequency at the nth harmonic is given by the orbital frequency of the binary $f_n=nf_{orb}$, and $f_{orb} = \sqrt{GM_{tot}/a^3}/2\pi$ where $M_{tot}$ is the total mass of the binary. The radiated power $\dot E_n$ can be written as \citep{Peters_Mathews_1963}
\begin{equation}
    \dot E_n = \frac{32}{5}\frac{G^{7/3}}{c^5}(2\pi f_{orb} \mathcal{M}_c)^{10/3}g(n,e),
\end{equation}
where $\mathcal{M}_c$ is the chirp mass at the rest frame and $\mathcal{M}_c = (M_{\rm{IMBH}}M_{\rm{WD}})^{3/5}/M_{tot}^{1/5}$. The time derivative of the frequency is computed to be
\begin{equation}
    \dot f_n = n\frac{96}{10\pi}\frac{(G\mathcal{M}_c)^{5/3}}{c^5}(2\pi f_{orb})^{11/3}F(e).
\end{equation}

\begin{figure}
\begin{center}
\includegraphics[width=\columnwidth]{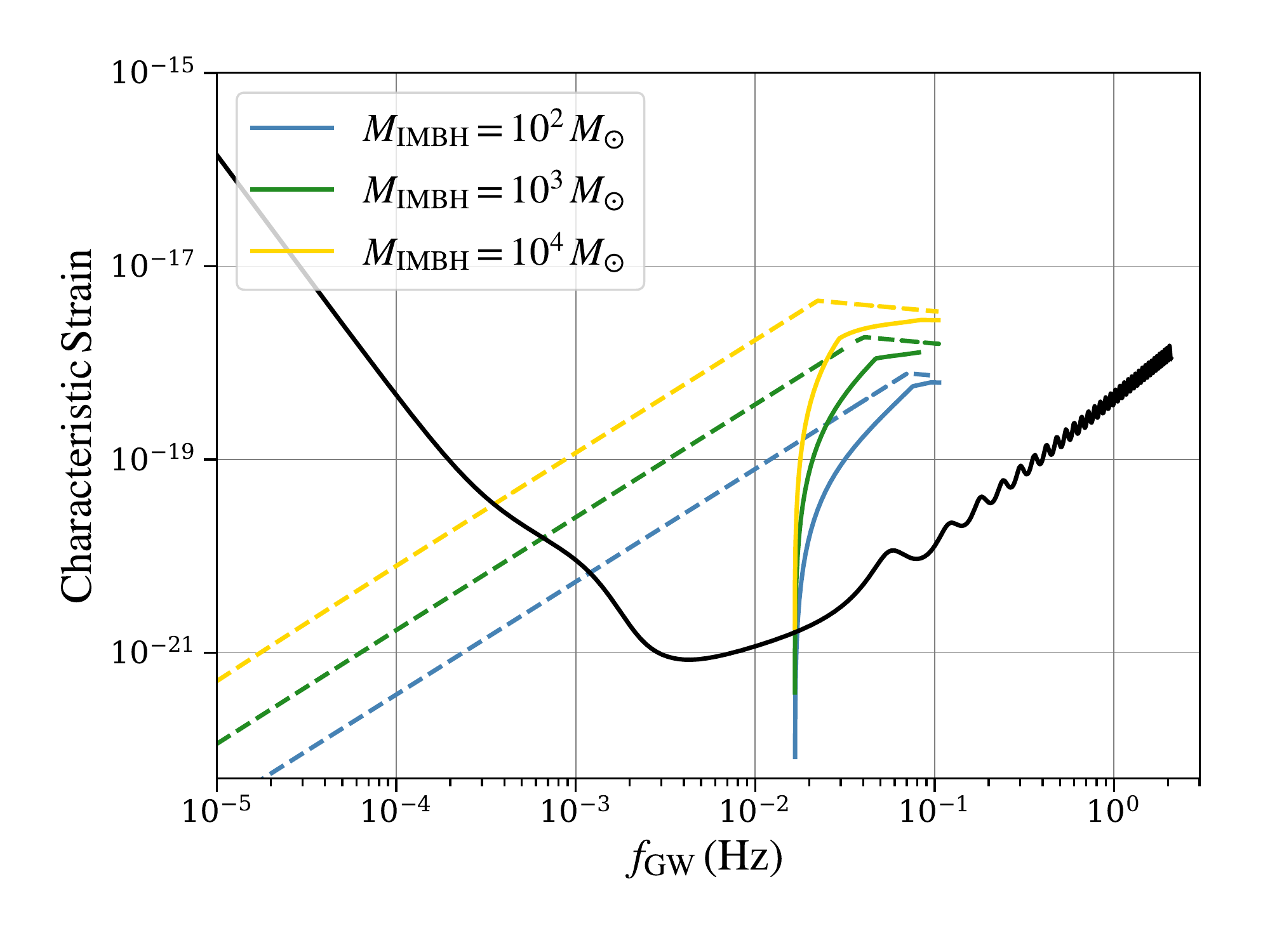}
\caption{Evolution of the characteristic strain at the peak frequency of GW emission assuming a distance of 1~Mpc. The evolution has initial $a=3R_{t}/(1-e)$ and $e=0.99$, and stops at the commencement of the tidal disruption of the WD. We assume LISA has a four-year mission lifetime. The dashed curves indicate the inspirals of circular IMBH-WD binaries for comparison.}\label{fig:gwstrain}
\end{center}
\end{figure}

Knowing the characteristic strains and the GW frequencies, we can calculate the signal-to-noise ratio S/N for eccentric binaries by summing over the relevant harmonics
\begin{equation}
    \left(\frac{S}{N}\right)^2 = \sum_{n=1}^{\infty} \int_{f_n}^{f^\prime_n} 
    \left[\frac{h_{c,n}}{h_f}\right]^2 \frac{d f_n}{f_n},
\end{equation}
where $h_f$ is the LISA noise curve from \citet{Robson+2019}, and $f_n^\prime$ is the GW frequency of the nth harmonic at the end of the LISA observation time. We truncate the calculation at the maximum harmonic \citep{OLeary+2009}
\begin{equation}
    n_{\rm{max}} = 5\frac{(1+e)^{0.5}}{(1-e)^{1.5}},
\end{equation} which is about $7053$ for $e=0.99$. Here we did not consider the orbital decay from tidal effects because \citet{Vick+2017} showed that they are negligible compared to GW radiation. The characteristic strains for binaries at 1~Mpc in comparison to the LISA sensitivity curve are shown as an example in Figure~\ref{fig:gwstrain}. Note that the characteristic strains are only accurate up to when tidal stripping starts since we assume no mass loss during the inspiral (\citealp{Zalamea+2010}; see Section~\ref{subsubsec:stripping} for more details).
The evolutionary time span of these strains (assuming inspiral stops at the tidal disruption radius) are about $390$, $1830$, and $8440$~years for $M_{\rm{IMBH}}=10^4$, $10^3$, and $10^2~\msun$, respectively. Combining the duration of the GW signals with the capture rates in Figure~\ref{fig:rate_nsc} and \ref{fig:rates_gc}, we can expect $\sim10^{-4}-10^{-3}$ events per Milky Way-like galaxy.

For S/N$>2$, the GW signals can be detected out to about $73$, $15$, and $3$~Mpc for $M_{\rm{IMBH}}=10^4$, $10^3$, and $10^2~\msun$, respectively. If we require that S/N$>10$, the maximum detectable distances are $14$, $3$, and $1$~Mpc instead for the same three IMBH masses. These values are consistent with \citet{Sesana+2008} which showed that the GW emission with S/N$>30$ can be detected out to $\approx 200$~Mpc for BHs with masses in the range of $\approx10^4-10^5\,\msun$, WDs with masses in the range of $\approx0.5-1\,\msun$, and circular orbits at tidal disruption or the innermost stable orbit.

\subsection{Electromagnetic Counterparts}\label{subsec:emcounter}

As discussed above and illustrated in Fig.~\ref{fig:times}, tidal capture dominates in GCs and at relatively larger distances in galactic nuclei, while in the innermost regions of the latter WDs are tidally disrupted more efficiently. The associated electromagnetic counterparts are expected to be qualitatively and quantitatively different in the two situations, and hence we discuss them separately below.

\subsubsection{Prompt Tidal Disruption (near parabolic orbit)}\label{subsubsec:tde}
Prompt tidal disruption of a WD by an IMBH will occur if the WD passes sufficiently close to the IMBH such that the pericenter $R_p$ of its orbit becomes smaller than the tidal disruption radius defined in Eq.~\ref{eq:Rt}. The strength of the encounter is measured by the penetration factor

\begin{equation}
\beta\equiv \frac{R_t}{R_p}
=\left(\frac{R_p}{R_{\rm{WD}}}\right)^{-1} \left(\frac{M_{\rm IMBH}}{M_{\rm{WD}}}\right)^{1/3}\,.
\label{eq:beta}
\end{equation}

The outcome of the encounter will depend on the value of $\beta$ \citep{Rosswog+2009}. Let us first consider the more typical case of $\beta \gtrsim 1$, which has been well studied in the limit in which the debris have a uniform spread in energy between a minimum value (the most bound debris) and a maximum (unbound) 
\citep{Rees_1988}. Under these conditions, about half of the debris from the tidal disruption remains bound to the IMBH and falls back, while the other half is unbound. Accretion of the bound debris onto the IMBH begins on a timescale determined by the fallback of the most bound debris

\begin{eqnarray}
&t_{\rm fb}& \approx 2\pi \left(
\frac{R_{\rm{WD}}^3 M_{\rm IMBH}}{G M_{\rm{WD}}^2}\right)^{1/2}
\approx 10^3~{\rm s}
\nonumber \\
&\times&
\left(\frac{M_{\rm IMBH}}{10^3 M_\odot}\right)^{1/2}
\left(\frac{M_{\rm{WD}}}{0.6 M_\odot}\right)^{-1}
\left(\frac{R_{\rm{WD}}}{10^4~{\rm km}}\right)^{3/2}.
\label{eq:fb}
 \end{eqnarray}
 
After an initial rapid rise, the late-time fallback rate (on timescales $>> t_{\rm fb}$) of the tidally disrupted bound debris is expected to follow a power-law decay as

\begin{equation}
\dot{M}_{\rm fb} (t) \approx  \dot{M}_{\rm fb} (t_{\rm fb}) \left(\frac{t}{t_{\rm fb}}\right)^{-5/3}\,,
\label{eq:mdot1}
\end{equation}

where the peak rate $\dot{M}_{\rm fb} (t_{\rm fb})$ is on the order of 

\begin{eqnarray}
&\dot{M}_{\rm fb}& (t_{\rm fb}) \approx \frac{M_{\rm{WD}}}{3t_{\rm fb}} \approx 2\times 10^{-4} \, M_\odot \,
{\rm s}^{-1} \nonumber \\
&\times&
\left(\frac{M_{\rm IMBH}}{10^3 M_\odot}\right)^{-1/2}
\left(\frac{M_{\rm{WD}}}{0.6 M_\odot}\right)^{2}
\left(\frac{R_{\rm{WD}}}{10^4~{\rm km}}\right)^{-3/2}.
\label{eq:dotm1peak}
\end{eqnarray}

The precise fate of the fallback material can only be determined via numerical simulations. Whether the fallback material is able to circularize depends on its ability to lose a significant amount of energy. Hydrodynamic simulations by \citet{Kawana+2018} show that WD-BH TDEs result in a variety of outcomes depending on the WD/BH masses and pericenter of the orbit. They find a subset of cases where self-interception of the stream occurs, favoring circularization of the debris and the onset of super-Eddington accretion rates. Strong outflows and relativistic jets are likely to ensue in these situations. The accretion rate onto the BH (and hence the emitted power) tracks the fallback rate as long as the viscous timescale of the disk is shorter than the fallback time. Assuming that a jet can be launched with efficiency $\epsilon_j$ and that a fraction $\epsilon_\gamma$ of the jet energy is dissipated into high energy radiation, then transients with luminosities $\sim (10^{46}-10^{47})(\epsilon_j/0.1)(\epsilon_\gamma/0.1) $~erg~s$^{-1}$ can be expected.

When the WD approaches the IMBH with a large penetration factor, it can be severely compressed to an extent that explosive nuclear burning can be ignited \citep{Luminet_Pichon_1989,Rosswog+2009}. During the period of compression, elements up to the iron group can be synthesized, which are then injected into the outflow with the explosion. The explosive transient would observationally manifest as a peculiar, underluminous thermonuclear explosion \citep{Rosswog+2008a}. At the same time, a sizable fraction of the debris ($\sim 35\%$, \citealt{Rosswog+2009}) would still remain bound and accrete to the BH, possibly after circularization onto a disk.

There have already been several unusual transients for which the tidal disruption of a WD by an IMBH provides a compelling explanation. \citet{Krolik2011} suggested this progenitor model for Swift~J1644+57 \citep{Burrows2011}, a bright X-ray source (peak isotropic luminosity $\sim 4 \times 10^{48}$~erg~s$^{-1}$) which, after a steady period of about 700~s, underwent several flaring events separated by longer timescales. These were interpreted as the result of repeated close passages and hence stripping. \citet{Shchrbakov2013} suggested that the unsual pair of the gamma-ray burst GRB060218 and the accompanying supernova SN2006aj may be the result of a WD which is tidally disrupted by a BH of $\sim 10^4 M_\odot$. The unusually long GRB, of about $ 2600$~s duration and equivalent isotropic luminosity $\sim  10^{47}$~erg~s$^{-1}$ \citep{Campana2006}, is compatible with the typical accretion timescale expected in these events, while the unusual supernova, the fastest of all the ones associated with GRBs, could have been the result of tidal pinching and ignition of the WD. More generally, tidal disruption of a WD by an IMBH is considered a good alternative model (to massive star collapse) for the subset of ultra-long GRBs (ULGRBs, \citealt{Levan+2014,Levan+2016}).
The rates of these events are rather uncertain, but a rough estimate is placed at about $0.01$~Myr$^{-1}$ per galaxy \citep{Gendre2013,Levan+2016}. Assuming (lacking direct observational data), that the rate of jetted TDEs from WDs disrupted by IMBHs is similar to that of the well studied events of stellar disruptions by SMBHs, and that is of a few percent of events \citep{vanvelzen2016}, then a rate of $0.1$~Myr$^{-1}$ as estimated in Section~3.1 for a typical case of a $10^3M_\odot$ IMBH and a $10^6 M_\odot$ SMBH  would imply a rate of $\sim 0.001$~Myr$^{-1}$ jetted events,
hence making WD disruptions by IMBH to be a potentially important channel of ULGB progenitors (i.e. $\sim 10\%$ with the estimates above).

Among transients of different nature, \citet{Peng2019} suggested that the tidal disruption of a WD by an IMBH is responsible for producing the two fast X-ray transients, CDF-S XT1 \citep{Bauer2017} and XT2 (\citealt{Xue2019}, with a peak luminosity $\sim 3\times 10^{45}$~erg~s$^{-1}$), characterized by an X-ray initial plateau lasting around hundreds to thousands of seconds, followed by a rapid decay in the light curve. Rates for these X-ray transients have been estimated to be in the range $\sim 1-10^3$~Gpc$^{-3}$ (depending on the redshift) from CDF-S XT1 \citep{Bauer2017}
and $\sim 10^4$~Gpc$^{-3}$ from CDF-S XT2
\citep{Xue2019}. Assuming a galaxy density of $10^{8}$~Gpc$^{-3}$ \citep{Conselice2016}, these estimates
would yield an event rate per galaxy for these transients in the range $10^{-2}-10^2$~Myr$^{-1}$. 
Since their X-ray emission has been modeled directly as the result of accretion (i.e. no jet requirement), our estimated disruption rates of 0.1~Myr$^{-1}$ would give plausibility to this interpretation for at least a fraction of transients.
Most recently, \citet{Gomez_Gezari_2023} has shown that SN~2020lrt may be a candidate for a WD being tidally compressed and disrupted by an IMBH.

\subsubsection{Tidal Stripping after Tidal Capture (eccentric orbit)} \label{subsubsec:stripping}

Once the WD has been captured by the IMBH, the orbit of the newly formed IMBH-WD binary will begin to shrink due to GW emission, as discussed in Section~\ref{subsec:gw}.

Mass loss in interacting binaries with eccentric orbits was studied by \citet{Sepinsky2007}. They showed that stripping of the outer layers of the WD begins when the pericenter of the orbit becomes small enough that the radius of the WD exceeds its Roche lobe, that is
\begin{equation}
R_{\rm{WD}} \gtrsim R_{\rm lobe}\approx \gamma R_p \left(\frac{M_{\rm IMBH}}{M_{\rm{WD}}}\right)^{-1/3}\,,
\label{eq:Rstrip}
\end{equation}
where the parameter $\gamma$ in the expression for the  Roche lobe depends on the orbital eccentricity, the mass ratio of the two objects in the binary, and the rotation of the WD. An average value is taken to be $\gamma\approx 0.5$. As the WD makes repeated passages through the pericenter of its orbit, mass gets stripped. This process was first studied semi-analytically by \citet{Zalamea+2010}, and subsequently, with increasing degree of sophistication, by \citet{MacLeod+2014}, \citet{Vick+2017} and \citet{Chen+2022}. The qualitative features of the phenomenon are similar in all these works, while quantitative details vary depending on the various approximations made. Here, we will follow the more recent \citet{Chen+2022}, who compared their analytical formalism against hydrodynamic simulations. For a penetration factor $\beta\lesssim 0.7$, they found that the fraction of mass lost, $\Delta M/M_{\rm{WD}}$, at each pericenter passage can be well approximated (when compared to numerical results) by the expression
\begin{equation}
\frac{\Delta M}{M_{\rm{WD}}} \simeq 4.8 \left[1-\left(\frac{M_{\rm{WD}}}{M_{\rm Ch}}\right)^{4/3}\right]^{3/4}\left(1-\frac{\gamma}{\beta}\right)^{5/2}\,,
\label{eq:delM}
\end{equation}
where $M_{\rm Ch}=1.4 M_\odot$ is the Chandrasekhar mass. The stripped debris is found to remain bound to the IMBH for eccentricities below a critical value
\begin{equation}
e_{\rm crit} \lesssim 1\,- \, 0.078\, \xi \,\beta \,M_{\rm IMBH}^{-1/3} \left(\frac{M_{\rm{WD}}}{0.6 M_\odot}\right)^{1/3}\,,
\label{eq:ecc_crit}
\end{equation}
where the parameter $\xi$ quantifies the uncertain tidal effects. From numerical simulations, they found $e_{\rm crit}\sim 0.8-0.9$ for a WD of $0.67 M_\odot$, and  $e_{\rm crit} \sim 0.7-0.8$ for a WD of $1.07 M_\odot$. 

\begin{figure}
\begin{center}
\includegraphics[width=\columnwidth]{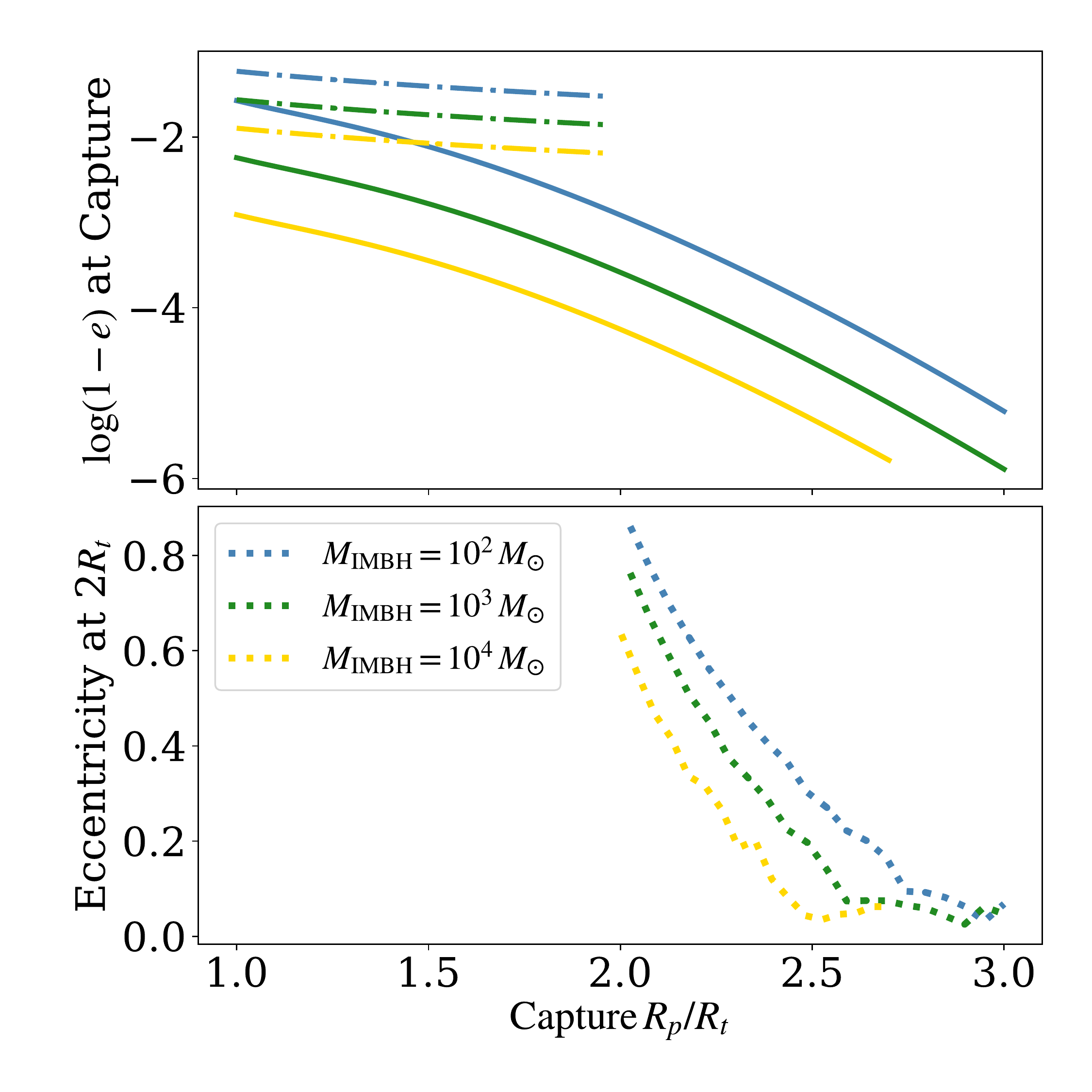}
\caption{Eccentricity distributions of the WDs at the moment of capture (upper panel, solid curves), and at the moment of initial mass loss after some degrees of inspiral (when $R_p\sim 2R_t$, bottom panel). Also plotted in the upper panel are the critical eccentricities for the stripped mass to remain bound to the IMBHs from Eq.~\ref{eq:ecc_crit} assuming $\xi=3.5$ (\citealp{Chen+2022}; dotted dash curves). The x-axis shows the pericenter distance at capture. The colors of the curves distinguish different IMBH masses.}
\label{fig:ecc}
\end{center}
\end{figure}

Fig.~\ref{fig:ecc} shows the distribution of eccentricities 
at capture (upper panel and solid lines), as well as at the moment that tidal stripping begins\footnote{Note that, if at the time that mass stripping starts the WD is rather found in a circular orbit, then the following evolution is characterized by stable mass transfer which, depending on the conditions, may lead to the WD to recede from the IMBH while filling its Roche lobe \citep{Dai2013}.} (bottom panel -- note that the condition in Eq.~\ref{eq:Rstrip} with $\gamma\approx 0.5$ is equivalent to $R_p\lesssim 2R_t$). While the eccentricity is close to one at capture, for systems that are captured with $R_p> 2R_t$, the eccentricity drops quickly by the time mass loss begins. Except for the narrow range of capture parameters $[2R_p/R_t-2.2 R_p/R_t]$, the eccentricity is below the critical value in Eq.~\ref{eq:ecc_crit}, and hence the stripped mass is expected to remain fully bound.

Once mass loss has begun, the  subsequent mass loss rate undergoes two phases. At early times, the mass loss follows the orbital shrinkage driven by the emission of GW radiation. Later, as the smaller WD expands and becomes less dense, it becomes easier to disrupt after each passage, until the WD is completely disrupted after a time $t_{\rm ML}$. \citet{Chen+2022} showed that this timescale is generally smaller than the GW timescale by a factor
\begin{equation}
\frac{t_{\rm ML}}{t_{\rm GW}}\simeq 0.02
\left(\frac{M_{\rm IMBH}}{10^4 M_\odot}\right)^{4/15}
\left(\frac{M_{\rm{WD}}}{0.6 M_\odot}\right)^{16/15}
\left(\frac{1-e_0}{0.1}\right)^{-1}\,,
\label{eq:tratio}
\end{equation}
where $e_0$ is the eccentricity at the beginning of the mass loss.

At the same time, mass loss will proceed with modulation on the binary period and increasing amplitude as the WD gets increasingly more stripped \citep{Zalamea+2010,MacLeod+2014}. The fallback rate of the tidally disrupted debris is also expected to display such a modulation, as long as the accretion timescale is smaller than the orbital period $P$. The calculations by \citet{Chen+2022} showed that a good approximation to the results of their numerical simulations for the ratio between these two timescales is
\begin{equation}
\frac{t_{\rm fb}}{P}\simeq \left[
1\,+\,\frac{2\xi\beta}{1-e}\left(
\frac{M_{\rm IMBH}}{M_{\rm{WD}}}\right)^{-1/3}
\right]^{-3/2}.
\label{eq:tfb}
\end{equation}
For the range of parameters $\beta$ (0.55-0.7), orbital parameters ($0.7\le e \le 0.9$ ), and WD masses (0.67 and 1.67 $M_\odot$) explored in their simulations, they found the above timescale ratio to be $\sim 0.2-0.5$.
The average fallback rate over an orbital period can be estimated as $\dot{M}_{\rm fb}\sim \Delta M/P$ (since, as discussed above, the greatest majority of our systems is found to be in a regime where all the stripped mass remains bound). For typical orbital parameters, its peak magnitude is found to exceed the Eddington value by several orders of magnitude, that is $\dot{M}_{\rm fb}\sim 10^{-5}-10^{-3}M_\odot$~s$^{-1}$ \citep{Zalamea+2010,MacLeod+2014,Chen+2022}. The modulation of the accretion rate onto the IMBH will further depend on the accretion (viscous) timescale if disk circularization occurs. The ratio between the accretion timescale and the orbital period can be estimated as (e.g. \citealt{MacLeod+2014})
\begin{equation}
\frac{t_{\rm acc}}{P} \simeq 0.2
\left(\frac{1-e}{0.03}\right)^{3/2}
\left(\frac{\alpha}{0.1}\right)^{-1}
\left(\frac{H/R}{0.5}\right)^{-2}\,,
\label{eq:tacc}
\end{equation}
in a viscously accreting ring of scale height $H$
with viscosity parameter $\alpha$ \citep{Shakura1973}. Given the range of $\alpha\sim 0.01-0.1$, these timescales can be comparable.  Modulation in the accretion rate onto the IMBH will still be expected even when partially smoothed out by viscosity.

Detailed predictions of the luminosity and its spectrum during the periodic mass loss phase are still lacking, since they depend heavily on whether an outflow and/or a jet can be driven. If the electromagnetic emission is the direct result of accretion onto the IMBH, then it will be Eddington-limited, and hence, for the IMBH masses considered here, luminosities would be limited to the $10^{40}-10^{42}$~erg~s$^{-1}$ range. However, considering that the accretion rates are highly super-Eddington, it is plausible that outflows and jets could be driven, in analogy to the standard TDE case discussed in the previous subsection. If a fraction $\epsilon (=\epsilon_j\epsilon_\gamma)$ of accreted mass is converted to luminosity, the peak fallback rates reported above would lead to an emission $L=\epsilon \dot{M_{\rm fb}} c^2 \sim 
(\epsilon/0.01) (10^{45}-10^{47}$)~erg~s$^{-1}$. The luminosity would be variable on the binary period and last for a timescale $\sim t_{\rm ML}$\footnote{Note that this timescale does not account for the modifications to the WD structure where the outer layer of the WD is expected to heat up and ignite before the point of disruption due to tidal heating  \citep{Vick+2017}.}.

Tidal stripping of a WD by an IMBH in a binary has already been invoked to explain some unusual, rare transients. \citet{Shen2019} suggested this progenitor system as 
potential progenitor candidate of the ultraluminous X-ray flares, with rapid rise ($\sim$ minute) and decay ($\sim $ hour) and peak luminosities $\sim 10^{40}-10^{41}$~erg~s$^{-1}$, that are observed in GCs and elliptical galaxies (\citealp{Irwin2016}; see also \citealt{Karpiuk2022} for modeling of these sources). Similarly, \citet{King2020} proposed the WD tidal stripping model to explain the quasi-periodic ultrasoft X-ray eruptions (with durations of $10^3-10^4$~sec, periods of about 9 hours, and peak luminosity $\sim 5\times 10^{42}$~erg~s$^{-1}$) observed from the galaxy GSN~069 \citep{Miniutti2019}. Sources of this brightness have been resolved by the {\em Chandra} telescope up to redshifts $\sim 4$ \citep{wilkes2019}, and will be detectable up to $\gtrsim 300$~Mpc with a short $\sim 10$~s exposure time by the Follow-up X-ray Telescope (FXT) of the upcoming mission {\em Einstein Probe} \citep{yuan2015}.

\section{Conclusions and Discussion} \label{sec:conclu}
In this study, we estimated the rates of WDs tidally captured by IMBHs in dense stellar environments. We assumed that a typical WD behaves as an $n=1.5$ polytrope and computed the capture radii based on the energy dissipated through WD oscillations during the first passage of the IMBH-WD encounter. We computed the capture rates for IMBHs with various masses in two environments, galactic nuclei and GCs. For the galactic nuclei, we adopted power-law distributions of stars and WDs around the central SMBHs, and studied how different SMBH masses and slopes of the stellar distribution affect the capture rates. We adopted Plummer profiles with different masses and half-mass radii in the case of GCs. Using these analytical models, we demonstrated that for a Milky Way-like galactic nucleus, the capture rate is $\sim 0.02$~per Myr for an IMBH of $10^3\,\msun$. Meanwhile, the capture rate is about a few times $0.1$ per Myr in a Milky Way-like galaxy if we assume that all GCs contain an IMBH. 

These captured IMBH-WD binaries are sources of intermediate-mass ratio inspirals and radiate GWs in the LISA frequency band. We calculated the characteristic strains and S/N for inspiralling eccentric IMBH-WD binaries and showed that their GW signals may be detected out to $\sim100$~Mpc by LISA. Following the phase of GW radiation which shrinks the orbits of the IMBH-WD binaries, tidal stripping and disruption of the WDs by the IMBHs will take effect. The luminosity of the electromagnetic emission during the periodic mass loss is at least $\gtrsim 10^{40}-10^{42}\,\rm{erg\,s^{-1}}$ (depending on the IMBH mass, if Eddington-limited). If aided by the launch of an outflow or jet, the luminosity may reach $\sim 10^{45}-10^{47}\,\rm{erg\,s^{-1}}$, comparable to that of standard TDEs \citep{MacLeod+2014}.

The capture and disruption rates and gravitational S/N estimated here should be taken as upper limits. We assumed that energy deposited into the WDs is efficiently dissipated, and the structure of the WDs stays unchanged during capture and inspiral. This essentially neglects any build-up of oscillation energy in the WDs and the possibility of WD disruption, allowing for all WDs that approach close to an IMBH to be captured into a lasting binary. For example, during very close approaches ($R_p \lesssim 2R_t$), the degeneracy of the WD's outermost layer may be lifted by shock heating, leading to faster mass loss \citep[e.g.,][]{Cheng_Evans_2013}. We also did not consider WD spins which could play an important role in the orbital evolution of the binaries \citep[e.g.,][]{Ivanov_Papaloizou_2007,Vick+2017}. These are reasonable assumptions for the first-order rate approximation here. A more detailed understanding of the WDs' reaction to the strong tidal force will require numerical calculations. For the computation of the GW emission, we followed the WD till its pericenter reaches the tidal radius of the IMBH. However, both the magnitude of the signal and its duration  should be considered as upper limits since tidal stripping reduces the WD mass and shortens the inspiral time. In addition, the IMBH occupation fraction and the mass distribution in galactic nuclei and in GCs are uncertain. Larger occupation fractions and flatter mass distributions will lead to higher tidal capture rates and vice versa. Future (non-)detection of IMBH-WD inspirals and electromagnetic emissions would be able to shed more light on the existence of IMBHs and potentially supply data for probing this wide range of physics.

\begin{acknowledgments}
We thank Fred Rasio, Kyle Kremer, Chris Matzner, and the anonymous referee for helpful discussions and comments. This work was supported by NSF Grants AST-1716762, AST-2108624 at Northwestern University, and by the Natural Sciences and Engineering Research Council of Canada (NSERC) DIS-2022-568580. G.F.\ acknowledges support from NASA Grant 80NSSC21K1722. 
R.P. acknowledges support by NSF award AST-2006839.
This research was supported in part through the computational resources and staff contributions provided for the Quest high performance computing facility at Northwestern University, which is jointly supported by the Office of the Provost, the Office for Research, and Northwestern University Information Technology.
\end{acknowledgments}

%\vspace{5mm}
%\facilities{HST(STIS), Swift(XRT and UVOT), AAVSO, CTIO:1.3m,


%\software{astropy \citep{2013A&A...558A..33A,2018AJ....156..123A},  
%          Cloudy \citep{2013RMxAA..49..137F}, 
%          Source Extractor \citep{1996A&AS..117..393B}
%          }


\bibliography{IMBHWD_TC}
\bibliographystyle{aasjournal}

\end{document}